\documentclass[12pts]{article}
\usepackage{CJK}
\usepackage{fancyhdr}
\usepackage{bm}
\usepackage{amsmath}
\usepackage{amssymb}
\usepackage{color}
\usepackage{graphicx}
\usepackage{subcaption}
\usepackage{multirow}
\usepackage{algorithm}
\usepackage{algorithmic} 
\usepackage{authblk}
\usepackage{hyperref}

\usepackage{graphicx}
\usepackage{setspace} 
\newcounter{ass}

\newcommand{\TT}{\mathcal{T}}

\usepackage{array} 
\title{Testing for cross-quantilogram change}
\author[$\dagger$]{Chia-Min Chang}
\author[$\dagger$]{Yu-Hsiang Cheng} 
\author[$\star$]{Tzee-Ming Huang\thanks{Corresponding Author: tmhuang@nccu.edu.tw}}
\affil[$\dagger$]{Department of Finance, Shih Hsin University, Taiwan (R.O.C.)}
\affil[$\star$]{Department of Statistics, National Chengchi Univeristy, Taiwan (R.O.C.)}
\begin{document}
\begin{CJK}{UTF8}{bkai}
\maketitle
\begin{abstract}
For two time series  $\{ (Y_t, \bm{Z}_t^Y) \}_{t\in \mathbb{Z}}$ and  $\{(X_t, \bm{Z}_t^X)\}_{t\in \mathbb{Z}}$, the directional dependence of $\{ X_t \}_{t \in \mathbb{Z}}$ on $\{ Y_t \}_{t\in\mathbb{Z}}$ while removing the impact of $\bm{Z}_t^X$ on $X_t$ and the impact of $\bm{Z}_t^Y$ on $ Y_t$ can be measured by cross-quantilograms. When the two time series are obeserved over two periods of time, it can be of interest to learn whether the cross-quantilograms remain the same for the two periods of time. We propose a test for this purpose, and the cross-quantilograms are estimated using the estimators proposed by \cite{han:2016}. The $p$-value of the proposed test is obtained based on a bootstrap approach. 
\end{abstract}

\begin{flushleft}
\end{flushleft}

\section{Introduction}
Suppose that $\{ (Y_t, \bm{Z}_t^Y) \}_{t\in \mathbb{Z}}$ and  $\{(X_t, \bm{Z}_t^X)\}_{t\in \mathbb{Z}}$ are two  multivariate time series that take values in $\mathbb{R}^{1+p}$ and $\mathbb{R}^{1+q}$ respectively,
where $\mathbb{Z}$ denotes the set of integers. Suppose that $\{ (Y_t, \bm{Z}_t^Y, X_t, \bm{Z}_t^X)\}_{t\in \mathbb{Z}}$ is stationary. To evaluate the directional dependence of $\{ X_t \}_{t\in \mathbb{Z}}$ on $\{ Y_t \}_{t\in \mathbb{Z}}$ while removing the impact of $\bm{Z}_t^X$ on $X_t$ and the impact of $\bm{Z}_t^Y$ on $ Y_t$,   Han et al. \cite{han:2016} proposed to use cross-quantilograms. 
 Specifically, given $\tau_1$, $\tau_2 \in (0,1)$ and a positive integer $k$,  the following cross-quantilogram $\rho_{\tau_1, \tau_2}(k)$ is considered in \cite{han:2016}:
\begin{equation}\label{def}
\rho_{\tau_1, \tau_2}(k) = \frac{E\Big(\psi_{\tau_1}\Big(Y_t - q_{Y, t}^{\tau_1}\Big)\psi_{\tau_2}\Big(X_{t-k} - q_{X, t-k}^{\tau_2}\Big)\Big)}{\sqrt{Var\Big(\psi_{\tau_1}\Big(Y_t - q_{Y, t}^{\tau_1}\Big)\Big)Var\Big(\psi_{\tau_2}\Big(X_{t-k} - q_{X, t-k}^{\tau_2}\Big)\Big)}}.
\end{equation}
Here $q_{Y, t}^{\tau_1}$ denotes the $\tau_1$ quantile of $Y_t$ conditional on $\bm{Z}_t^Y$, $q_{X, t-k}^{\tau_2}$ denotes the $\tau_2$ quantile of $X_{t-k}$ conditional on $\bm{Z}_{t-k}^X$, and  for  a constant $\tau \in (0,1)$, the function $\psi_\tau$ is defined by
\[
 \psi_\tau(x)  = \left\{ \begin{array}{ll} 
 1 - \tau & \mbox{ if } x < 0; \\
 -\tau & \mbox{ if } x \geq 0, \\
 \end{array} \right. 
\]
for $x \in \mathbb{R}$.

In \cite{han:2016}, one problem of interest is to test whether $\rho_{\tau_1, \tau_2}(k) = 0$ for every $(\tau_1, \tau_2, k) \in \TT_1 \times \TT_2 \times \{ 1, \ldots, p \}$, where  $\TT_1$, $\TT_2$ are pre-specified subsets of $(0,1)$ and $p$ is a  pre-specified positive integer. In this study,  our problem of interest is different.
We assume that the series $\{ (Y_t, \bm{Z}_t^Y, X_t, \bm{Z}_t^X)\}_{t\in Z}$ is observed over two periods of time, $\{ (Y_{b, t}, \bm{Z}_{b, t}^Y, X_{b, t}, \bm{Z}_{b, t}^X) \}_{t=1}^{T_1}$ and $\{ (Y_{a, t}, \bm{Z}_{a, t}^Y, X_{a, t}, \bm{Z}_{a, t}^X) \}_{t=1}^{T_2}$, and we are interested in examining whether the cross-quantilograms are the same across these two periods. This issue can arise naturally when anaylzing data in economics and finance. For instance, one may be interested in examining whether the directional dependence between two assets changes when a specific event occurs. 

The rest of this paper is organized as follows. In Section 2, the proposed testing method is presented. In Section 3, simulation results are showed to demonstrate the performance of the proposed test. In Section 4, an application of the proposed method to real financial data is given. Finally, conclusions regarding the method are provided.

\section{Method} 
As mentioned before,  we are interested in testing  whether the cross-quantilograms are the same across these two periods. The cross-quantilogram estimation in our proposed test is based on the estimator in (\ref{def}), so we first review the  cross-quantilogram estimator in  (\ref{def}) before stating our test. 

For the cross-quantilogram $\rho_{\tau_1, \tau_2}(k)$ in (\ref{def}), the proposed estimator of $\rho_{\tau_1, \tau_2}(k)$ in \cite{han:2016} based on 
observations  $\{ (Y_t, \bm{Z}_t^Y) \}_{t=1}^T$ and  $\{(X_t, \bm{Z}_t^X)\}_{t=1}^T$ is given by
\begin{equation}\label{sample}
\widehat{\rho}_{\tau_1, \tau_2}(k) = \frac{\sum_{t=k+1}^{T}\psi_{\tau_1}\Big(Y_t - \hat{q}_{Y, t}^{\tau_1}\Big)\psi_{\tau_2}\Big(X_{t-k} - \hat{q}_{X, t-k}^{\tau_2}\Big)}{\sqrt{\sum_{t=k+1}^T\Big(\psi_{\tau_1}\Big(Y_t - \hat{q}_{Y, t}^{\tau_1}\Big)\Big)^2}\sqrt{\sum_{t=k+1}^T\Big(\psi_{\tau_2}\Big(X_{t-k} - \hat{q}_{X, t-k}^{\tau_2}\Big)\Big)^2}}.
\end{equation}
Here, $\hat{q}_{Y, t}^{\tau_1}$ and $\hat{q}_{X, t-k}^{\tau_2}$ are estimators of $q_{Y, t}^{\tau_1}$ and $q_{X, t-k}^{\tau_2}$ based on $\{ (Y_t, \bm{Z}_t^Y) \}_{t=1}^T$ and  $\{(X_t, \bm{Z}_t^X)\}_{t=1}^T$, respectively.  The estimator $\hat{q}_{Y, t}^{\tau_1}$ is obtained by first fitting the linear quantile regression model proposed by Koenker and Bassett \cite{Koenker:1978} to the data $\{ (Y_t, \bm{Z}_t^Y) \}_{t=1}^T$,  using $\bm{Z}_1^Y$, $\ldots$, $\bm{Z}_T^Y$ as the observations of the predictors and $Y_1$, $\ldots$, $Y_T$ as the corresponding response observations, and then computing $\hat{q}_{Y, t}^{\tau_1}$ as the resulting estimated conditional quantile of the response when the predictor is $\bm{Z}_t^Y$.  When there is no predictor in the linear quantile regression model, $\hat{q}_{Y, t}^{\tau_1}$ is the sample quantile on $\{Y_1, \ldots, Y_T\}$. 
The estimator $\hat{q}_{X, t-k}^{\tau_2}$ is obtained in a similar way to that in which  $\hat{q}_{Y, t}^{\tau_1}$ is obtained, with $\{ (Y_t, \bm{Z}_t^Y) \}_{t=1}^T$  replaced by  $\{(X_t, \bm{Z}_t^X)\}_{t=1}^{T-k}$. 

In \cite{han:2016}, in order to examime whether $\rho_{\tau_1, \tau_2}(k) = 0$ for every $(\tau_1, \tau_2, k) \in \TT_1 \times \TT_2 \times \{ 1, \ldots, p \}$, where  $\TT_1 \times \TT_2$ is a pre-specified subset of $\Omega$ and $p$ is a  pre-specified positive integer, the test statistic
\[\sup_{(\tau_1, \tau_2) \in \Omega}T\sum_{k=1}^p \widehat{\rho}^2_{\tau_1, \tau_2}(k)\]
was proposed in \cite{han:2016}. Moreover, in \cite{han:2016}, certain asymptotic properties of the test statistic were established and an approach for finding the $p$-value for the test is proposed based on the derived asymptotic properties and a stationary bootstrap resampling method proposed by  K\"unsch \cite{Kunsch:1989}.

We are now ready to describe our test. As mentioned in Section 1, we observe two  series $\{ (Y_{b, t}, \bm{Z}_{b, t}^Y, X_{b, t}, \bm{Z}_{b, t}^X) \}_{t=1}^{T_1}$ and $\{ (Y_{a, t}, \bm{Z}_{a, t}^Y, X_{a, t}, \bm{Z}_{a, t}^X) \}_{t=1}^{T_2}$ that are independent of each other, and  it is of our interest to test whether
\begin{equation} \label{eq:change}
 H_0: \rho_{b, \tau_1, \tau_2}(k)  = \rho_{a, \tau_1, \tau_2}(k) \mbox{ for all $(\tau_1, \tau_2, p) \in \TT_1\times \TT_2 \times \{ 1, \ldots p\}$, } 
\end{equation}
where $\rho_{b, \tau_1, \tau_2}(k)$ and $\rho_{a, \tau_1, \tau_2}(k)$ denote the cross-quantilogram $\rho_{\tau_1, \tau_2}(k)$ defined in \ref{def} with  $\{ (Y_t, \bm{Z}_t^Y, X_t, \bm{Z}_t^X)\}_{t\in Z}$ replaced by $\{ (Y_{b, t}, \bm{Z}_{b, t}^Y, X_{b, t}, \bm{Z}_{b, t}^X) \}_{t=1}^{T_1}$ and $\{ (Y_{a, t}, \bm{Z}_{a, t}^Y, X_{a, t}, \bm{Z}_{a, t}^X) \}_{t=1}^{T_2}$, respectively.  Let $\widehat{\rho}_{b, \tau_1, \tau_2}(k)$ and $\widehat{\rho}_{a, \tau_1, \tau_2}(k)$ denote the estimator $\widehat{\rho}_{\tau_1, \tau_2}(k)$ in (\ref{sample}) with 
 $\{ (Y_t, \bm{Z}_t^Y, X_t, \bm{Z}_t^X)\}_{t=1}^T$ replaced by $\{ (Y_{b, t}, \bm{Z}_{b, t}^Y, X_{b, t}, \bm{Z}_{b, t}^X) \}_{t=1}^{T_1}$ and $\{ (Y_{a, t}, \bm{Z}_{a, t}^Y, X_{a, t}, \bm{Z}_{a, t}^X) \}_{t=1}^{T_2}$, respectively. 
Let $\Omega=\TT_1 \times \TT_2$, then our proposed test  for (\ref{eq:change}) is based on the test statistic
\begin{equation} \label{eq:D_estimator}
\hat{D} = \sup_{\tau_1, \tau_2 \in \Omega} \sum_{k=1}^p \left( \widehat{\rho}_{b, \tau_1, \tau_2}(k)  - \widehat{\rho}_{a, \tau_1, \tau_2}(k) \right)^2.
\end{equation}
The null hypothesis $H_0$ in (\ref{eq:change}) is rejected if the test statistic $\hat{D}$ is large. To simplify the notation, we will denote the lag $k$ cross-quantilogram estimators $\widehat{\rho}_{b, \tau_1, \tau_2}(k) $ and $ \widehat{\rho}_{a, \tau_1, \tau_2}(k)$ by $\widehat{\rho}_b(k)$ and $\widehat{\rho}_a(k)$, respectively.

To obtain the $p$-value of the proposed test, we follow the steps below:
\begin{itemize}
\item[(i)] Compute the lag $k$ cross-quantilogram estimators  $\widehat{\rho}_b(k)$ and $\widehat{\rho}_a(k)$ based on the original data  $\{ (Y_{b, t}, \bm{Z}_{b, t}^Y, X_{b, t}, \bm{Z}_{b, t}^X) \}_{t=1}^{T_1}$ and $\{ (Y_{a, t}, \bm{Z}_{a, t}^Y, X_{a, t}, \bm{Z}_{a, t}^X) \}_{t=1}^{T_2}$ for $k=1$, $\ldots$, $p$.

\item[(ii)] Generate bootstrapping data $Data_1$ and $Data_2$ independently from the original data $\{ (Y_{b, t}, \bm{Z}_{b, t}^Y, X_{b, t}, \bm{Z}_{b, t}^X) \}_{t=1}^{T_1}$ and $\{ (Y_{a, t}, \bm{Z}_{a, t}^Y, X_{a, t}, \bm{Z}_{a, t}^X) \}_{t=1}^{T_2}$ respectively following the approach in \cite{han:2016}, and for $k=1$, $\ldots$, $p$, let $\widehat{\rho}_{b, B}(k)$ and $\widehat{\rho}_{a, B}(k)$ denote the lag $k$ cross-quantilogram estimators based on $Data_1$ and $Data_2$, respectively. 

\item[(iii)] Repeat (ii) $L$ times and let $(\widehat{\rho}_{b, B, \ell}(k), \widehat{\rho}_{a, B, \ell}(k))$ denote the $(\widehat{\rho}_{b, B}(k),\widehat{\rho}_{a, B}(k))$ for $k=1$, $\ldots$, $p$ from the $\ell$-th repetition of (ii) for $\ell \in \{ 1, \ldots, L \}$.
Let 
\[ \hat{D} = \sup_{\tau_1, \tau_2 \in \Omega} \sum_{k=1}^p \Big(\widehat{\rho}_b(k)-\widehat{\rho}_a(k)\Big)^2\]
and let
\[
 \hat{D}_{B,\ell} =  \sup_{\tau_1, \tau_2 \in \Omega} \sum_{k=1}^p \Big(\big(\widehat{\rho}_{b, B, \ell}(k)-\widehat{\rho}_b(k)\big)-\big(\widehat{\rho}_{a, B, \ell}(k)-\widehat{\rho}_a(k)\big)\Big)^2
\]
for $\ell \in \{ 1, \ldots, L \}$.

\item[(iv)] The $p$-value of the proposed test is
\[
  \frac{\sum_{\ell=1}^{L} I(\hat{D}  <  \hat{D}_{B,\ell} )}{L},
\]
where 
\[
 I(\hat{D}  <  \hat{D}_{B,\ell} ) = \left\{ \begin{array}{ll} 
 1 & \mbox{ if } \hat{D}  <  \hat{D}_{B,\ell}; \\
 0 & \mbox{ if } \hat{D}  \geq  \hat{D}_{B,\ell}. \\
 \end{array} \right.
\]
\end{itemize}

The above process for obtaining the $p$-value of the proposed test can be justified based on the asymptotic results in \cite{han:2016} and the assumption that the two series $\{ (Y_{b, t}, \bm{Z}_{b, t}^Y, X_{b, t}, \bm{Z}_{b, t}^X) \}_{t=1}^{T_1}$ and $\{ (Y_{a, t}, \bm{Z}_{a, t}^Y, X_{a, t}, \bm{Z}_{a, t}^X) \}_{t=1}^{T_2}$ are independent and each series is stationary. To provide theoretic justification, one would need to make techinical assumptions like those in \cite{han:2016}. In this paper, we focus on evaluating the performance of the proposed test via simulation study. The details of our simulation study will be presented in the next section.
 
\section{Simulation study}
To investigate the performance of the proposed test, we conducted a simultion study with three experiments. In each experiment, we generated 
two time series  $\{ (Y_{b,t}, \bm{Z}_{b,t}^Y, X_{b,t}, \bm{Z}_{b,t}^X) \}_{t=1}^{T_1}$ and $\{ (Y_{a,t}, \bm{Z}_{a,t}^Y, X_{a,t}, \bm{Z}_{a,t}^X) \}_{t=1}^{T_2}$ independently and then performed the proposed test to examine its power.  

Below we first give the details of our data generating processes in the three experiments. In each experiment,  $\{ (Y_{b,t}, \bm{Z}_{b,t}^Y, X_{b,t}, \bm{Z}_{b,t}^X) \}_{t=1}^{T_1}$ and $\{ (Y_{a,t}, \bm{Z}_{a,t}^Y, X_{a,t}, \bm{Z}_{a,t}^X) \}_{t=1}^{T_2}$ were according to the distribution of $\{ (Y_{t}, \bm{Z}_{t}^Y, X_{t}, \bm{Z}_{t}^X ) \}_t$, where $\{ (X_t, Y_t) \}_{t=1}^T$ was generating according to one of the two models P1 and P2 stated below:
\begin{itemize}
	\item[P1:]
	\begin{align*}
		Y_t &= 0.15 + \alpha_0 Y_{t-1} + \alpha_1 X_{t-1} + \epsilon_{1,t}\hspace{8cm}\\
		X_t &=0.05 + \alpha_0 X_{t-1} + \epsilon_{2,t}, 
	\end{align*}
	where $\epsilon_{1,t}$ and $\epsilon_{2,t}$ are independent random variables from the standard normal distribution.  
	\item[P2:]
	\begin{align*}
		Y_t &= \beta_0 + \beta_1 Y_{t-1} +  U_t\hspace{8.5cm}\\
		X_t &= \gamma_0 + \gamma_1 X_{t-1} + V_t, 
	\end{align*}
	where $\{ (U_t,V_t) \}_t$ are generated according to the distribution of  $\{ (Y_t, X_t) \}_t$ in P1. 
\end{itemize}
When generating data from P1 or P2  in the three experiments,  we removed the first 5000 observations so that the remaining series  behaved  like a stationary bivariate time series. We took the remaining series as $\{ (X_t, Y_t) \}_{t=1}^T$.

For the specification of controlling variables $\bm{Z}_{t}^Y$ and $\bm{Z}_{t}^X$ based on $\{ (X_t, Y_t) \}_{t=1}^T$, we considered taking a controlling variable to be a lagged variable of $X_t$ or $Y_t$ such as $X_{t-1}$, $Y_{t-1}$, or not using controlling variable. In the case where the controlling variable $\bm{Z}_{t}^Y$ (or $\bm{Z}_{t}^X$) is not used, we will denote $\bm{Z}_{t}^Y$ (or $\bm{Z}_{t}^X$) by None.

The model choice (P1 or P2) and the specification of controlling variables for  $\{ (Y_{b,t}, \bm{Z}_{b,t}^Y, X_{b,t}, \bm{Z}_{b,t}^X) \}_{t=1}^{T_1}$ and $\{ (Y_{a,t}, \bm{Z}_{a,t}^Y, X_{a,t}, \bm{Z}_{a,t}^X) \}_{t=1}^{T_2}$ was according to the distribution of $\{ (Y_{t}, \bm{Z}_{t}^Y, X_{t}, \bm{Z}_{t}^X ) \}_t$ are the same, but the model parameters can be different. The model parameter vectors $(\alpha_0$, $\alpha_1$, $\beta_0$, $\beta_1)$ for 
$\{ (Y_{b,t}, \bm{Z}_{b,t}^Y, X_{b,t}, \bm{Z}_{b,t}^X) \}_{t=1}^{T_1}$ and $\{ (Y_{a,t}, \bm{Z}_{a,t}^Y, X_{a,t}, \bm{Z}_{a,t}^X) \}_{t=1}^{T_2}$  are denoted by $(\alpha_0^b$, $\alpha_1^b$, $\beta_0^b$, $\beta_1^b)$ and $(\alpha_0^a$, $\alpha_1^a$, $\beta_0^a$, $\beta_1^a)$, respectively.
The model choice and the specification of controlling variables $\bm{Z}_{t}^Y$ and $\bm{Z}_{t}^X$ in the three experiments are stated below.
\begin{itemize}
	\item Experiment 1
	\begin{itemize}
		\item Model: P1
		\item $(Y_t, \bm{Z}_t^Y) = (Y_t, None), (X_t, \bm{Z}_t^X) = (X_t, None)$
	\end{itemize}
	\item Experiment 2
	\begin{itemize}
		\item Model: P1
		\item $(Y_t, \bm{Z}_t^Y) = (Y_t, (Y_{t-1}, X_{t-1})), (X_t, \bm{Z}_t^X) = (X_t, X_{t-1})$
	\end{itemize}
	\item Experiment 3
	\begin{itemize}
		\item Model: P2
		\item $(Y_t, \bm{Z}_t^Y) = (Y_t, Y_{t-1}), (X_t, \bm{Z}_t^X) = (X_t, X_{t-1})$
	\end{itemize}
\end{itemize}

In these three experiments, the sample sizes of the generated time series, $T_1$ and $T_2$, were set to 500 or 1,000. Moreover, we set $T_1 = T_2$ for simplicity. The details of model coefficients are presented in Table 1 and 2. In addition, both the ranges of $\tau_1$ and $\tau_2$ ($\TT_1$ and $\TT_2$) were taken to be 
$\TT = \{0.05, 0.1, \ldots, 0.95\}$ and the bootstrap sample size $L$ was set to 800.  

Tables 1 and 2 present the performance of Experiments 1, 2, and 3. The reported power values are power estimates that are frequencies of rejecting the null hypothesis 
\[\sup_{(\tau_1, \tau_2) \in \TT^2 }\sum_{k=1}^5\Big(\rho_b(k)-\rho_a(k)\Big)^2 = 0\]
among 1000 trials divided by 1000.

In order to present the strength of the directional dependence between the two time series in each case, we included the "Difference" column in the tables, which gives approximate values of 
\[
 D = \sup_{(\tau_1, \tau_2) \in \TT^2}\sum_{k=1}^5\Big(\rho_b(k)-\rho_a(k)\Big)^2,
\]
where $\rho_b(k)$ and $\rho_a(k)$ denote the cross-quantilogram at lag $k$ for $\{ (Y_{b,t}, X_{b,t}) \}_{t=1}^{T_1}$ and $\{ (Y_{a,t}, X_{a,t}) \}_{t=1}^{T_2}$, respectively. Each approximate value of $D$ was obtained by generating $\{ (Y_{b,t}, X_{b,t}) \}_{t=1}^{T_1}$ and $\{ (Y_{a,t}, X_{a,t}) \}_{t=1}^{T_2}$ for very large $T_1$ and $T_2$ 20 times, computing the estimated cross-quantilogram in each trial, and then computing the averages of the 20 cross-quantilogram estimates as an approximate value of $D$.

The results of Experiment 1 are presented in Table 1. The approximate $D$ values are quite different, indicating different levels of strength in directional dependence between the two time series. 

From the results in Table 1, it can be found that, for each experiment at a given sample size, the power of the proposed test increases as the value $D$ increases. and when 
the value of $D$ is large enough, the power of the proposed test is satisfactory. Moreover, as the sample size increases, the test power improves significantly. The results in Experiment 3 show a similar pattern. 

When $D\approx 0$, there is almost no difference in directional dependence between the two time series, The Type I error of the proposed test is properly controlled. This can be found from the results of Experiment 2 and the last case of Experiment 1 in Table 1 and the last two cases of Experiment 3 in Table 2.

\begin{table}[h]
\tabcolsep=0.195cm
\renewcommand\arraystretch{1.3} 
\setlength{\abovecaptionskip}{-0.2cm}
\caption{The results for Experiments 1 and 2 }
\begin{center}
\begin{tabular}{ccccccc}
\hline
\hline
$(\alpha_0^b, \alpha_1^b)$ & $(\alpha_0^a, \alpha_1^a)$  & Sample size  &  \multicolumn{2} {c}{Experiment 1} &  \multicolumn{2} {c}{Experiment 2}\\
& & & Difference & Power & Difference & Power\\
\hline 
\multirow{2}{*}{(0.5, -0.4)} & \multirow{2}{*}{(0.5, -0.1)} & 500 & \multirow{2}{*}{0.111} & 0.665 & \multirow{2}{*}{0.00032} & 0.033\\
& &  1000 &  & 0.964 & & 0.046\\
\multirow{2}{*}{(0.5, -0.4)} & \multirow{2}{*}{(0.5, 0)} & 500 & \multirow{2}{*}{0.214} & 0.963 & \multirow{2}{*}{0.00031} &0.035\\
& &  1000 & & 1 & & 0.046\\
\multirow{2}{*}{(0.5, -0.4)} & \multirow{2}{*}{(0.5, 0.1)}  & 500 & \multirow{2}{*}{0.350} & 0.997 & \multirow{2}{*}{0.00031} & 0.035\\
& &  1000 & & 1 & & 0.039\\
\multirow{2}{*}{(0.5, -0.4)} & \multirow{2}{*}{(0.5, 0.4)} & 500 & \multirow{2}{*}{0.848} & 1 & \multirow{2}{*}{0.00029} & 0.034\\
& &  1000 & & 1 & & 0.043\\
\multirow{2}{*}{(0.5, -0.1)} & \multirow{2}{*}{(0.5, 0)} & 500 & \multirow{2}{*}{0.017} & 0.115 &\multirow{2}{*}{0.00029} & 0.026\\
& &  1000 & & 0.240 & & 0.040\\
\multirow{2}{*}{(0.5, -0.1)} & \multirow{2}{*}{(0.5, 0.1)}  & 500 & \multirow{2}{*}{0.067}& 0.459 & \multirow{2}{*}{0.00030}& 0.029\\
& &  1000 & & 0.848 & & 0.038\\
\multirow{2}{*}{(0.5, -0.1)} & \multirow{2}{*}{(0.5, 0.4)} & 500 & \multirow{2}{*}{0.350} & 0.998 &\multirow{2}{*}{0.00031} & 0.034\\
& &  1000 & & 1 & & 0.040\\
\multirow{2}{*}{(0.5, 0)} & \multirow{2}{*}{(0.5, 0.1)} & 500 & \multirow{2}{*}{0.017} & 0.103 & \multirow{2}{*}{0.00030}& 0.036\\
& &  1000 & & 0.257 & & 0.047\\
\multirow{2}{*}{(0.5, 0)} & \multirow{2}{*}{(0.5, 0.4)} & 500 & \multirow{2}{*}{0.214} & 0.956 &\multirow{2}{*}{0.00032} & 0.036\\
& &  1000 & & 1 & & 0.042\\
\multirow{2}{*}{(0.5, 0.1)} & \multirow{2}{*}{(0.5, 0.4)} & 500 & \multirow{2}{*}{0.111} & 0.665 & \multirow{2}{*}{0.00031}& 0.036\\
&  & 1000 & & 0.997 & & 0.040\\
\multirow{2}{*}{(-0.5, 0.4)} & \multirow{2}{*}{(0.2, 0.4)} & 500 & \multirow{2}{*}{0.176} & 0.982 & \multirow{2}{*}{0.00030}& 0.036\\
& &  1000 & & 1 & & 0.043\\
\multirow{2}{*}{(-0.5, 0.4)} & \multirow{2}{*}{(-0.5, 0.4)} & 500 & \multirow{2}{*}{$<$0.0001} & 0.025 & \multirow{2}{*}{0.00030}& 0.035\\
& &  1000 & & 0.037 & & 0.043\\
\hline
\hline
\end{tabular}
\label{table:1}
\end{center}
\end{table}

\begin{table}[h]
\tabcolsep=0.18cm
\renewcommand\arraystretch{1.3} 
\setlength{\abovecaptionskip}{-0.2cm}
\caption{The results for Experiment 3}
\begin{center}
\setlength{\tabcolsep}{1.85pt}
\begin{tabular}{ccccc}
\hline
\hline
$(\beta_0^b, \beta_1^b, \gamma_0^b, \gamma_1^b, \alpha_0^b, \alpha_1^b)$ & $(\beta_0^a, \beta_1^a, \gamma_0^a, \gamma_1^a, \alpha_0^a, \alpha_1^a)$  & Difference & Sample size & Power \\ 
\multirow{2}{*}{(0.1, 0.2, -0.1, -0.3, 0.2, -0.4)} & \multirow{2}{*}{(0.1, 0.2, -0.1, -0.3, 0.2, -0.1)} & \multirow{2}{*}{0.033} & 500 & 0.602 \\
& & & 1000 & 0.960 \\
\multirow{2}{*}{(0.1, 0.2, -0.1, -0.3, 0.2, -0.4)} & \multirow{2}{*}{(0.1, 0.2, -0.1, -0.3, 0.2, 0)} & \multirow{2}{*}{0.061} & 500 & 0.916 \\
& & & 1000 & 1 \\
\multirow{2}{*}{(0.1, 0.2, -0.1, -0.3, 0.2, -0.4)} & \multirow{2}{*}{(0.1, 0.2, -0.1, -0.3, 0.2, 0.1)} & \multirow{2}{*}{0.096} & 500 & 0.996 \\
& & & 1000 & 1 \\
\multirow{2}{*}{(0.1, 0.2, -0.1, -0.3, 0.2, -0.4)} & \multirow{2}{*}{(0.1, 0.2, -0.1, -0.3, 0.2, 0.4)} & \multirow{2}{*}{0.239} & 500 & 1 \\
& & & 1000 & 1 \\
\multirow{2}{*}{(0.1, 0.2, -0.1, -0.3, 0.2, -0.1)} & \multirow{2}{*}{(0.1, 0.2, -0.1, -0.3, 0.2, 0)} & \multirow{2}{*}{0.005} & 500 & 0.078 \\
& & & 1000 & 0.131 \\
\multirow{2}{*}{(0.1, 0.2, -0.1, -0.3, 0.2, -0.1)} & \multirow{2}{*}{(0.1, 0.2, -0.1, -0.3, 0.2, 0.1)} & \multirow{2}{*}{0.017} & 500 & 0.257 \\
& & & 1000 & 0.644 \\
\multirow{2}{*}{(0.1, 0.2, -0.1, -0.3, 0.2, -0.1)} & \multirow{2}{*}{(0.1, 0.2, -0.1, -0.3, 0.2, 0.4)} & \multirow{2}{*}{0.097} & 500 & 0.992 \\
& & & 1000 & 1 \\
\multirow{2}{*}{(0.1, 0.2, -0.1, -0.3, 0.2, 0)} & \multirow{2}{*}{(0.1, 0.2, -0.1, -0.3, 0.2, 0.1)} & \multirow{2}{*}{0.005} & 500 & 0.072 \\
& & & 1000 & 0.151 \\
\multirow{2}{*}{(0.1, 0.2, -0.1, -0.3, 0.2, 0)} & \multirow{2}{*}{(0.1, 0.2, -0.1, -0.3, 0.2, 0.4)} & \multirow{2}{*}{0.061} & 500 & 0.917 \\
& & & 1000 & 1 \\
\multirow{2}{*}{(0.1, 0.2, -0.1, -0.3, 0.2, 0.1)} & \multirow{2}{*}{(0.1, 0.2, -0.1, -0.3, 0.2, 0.4)} & \multirow{2}{*}{0.033} & 500 & 0.571 \\
& & & 1000 & 0.955 \\
\multirow{2}{*}{(-0.1, -0.2, -0.1, -0.3, 0.2, -0.4)} & \multirow{2}{*}{(0.1, 0.2, -0.1, -0.3, 0.2, 0.1)} & \multirow{2}{*}{0.098} & 500 & 0.997 \\
& & & 1000 & 1 \\
\multirow{2}{*}{(0.1, 0.2, -0.1, -0.3, 0.2, 0.1)} & \multirow{2}{*}{(0.1, 0.2, -0.1, -0.3, 0.2, 0.1)} & \multirow{2}{*}{0.0003} & 500 & 0.028 \\
& & & 1000 & 0.047 \\
\multirow{2}{*}{(-0.1, -0.2, -0.1, -0.3, 0.2, 0.1)} & \multirow{2}{*}{(0.1, 0.2, -0.1, -0.3, 0.2, 0.1)} & \multirow{2}{*}{0.0003} & 500 & 0.039 \\
& & & 1000 & 0.057 \\
\hline
\hline
\end{tabular}
\label{table:1}
\end{center}
\end{table}
\section{A analysis of OVX and XAU dataset}
In this section, we presented the results of applying the proposed test to a dataset that includes CBOE Crude Oil Volatility Index (OVX) and XAU/USD (XAU) data to examine whether directional dependence between the two series changed after the Russia-Ukraine War. OVX is calculated by the Chicago Board Options Exchange based on option prices of the United States Oil Fund (USO) and is a measure of the expected future volatility of crude oil prices. Note that XAU usually represents the spot price of gold, which is quoted in U.S. dollars, but in this analysis, we worked with the differenced price series, which we denoted by XAU or XAU returns.

The dataset used in this study contains OVX data and XAU data from December 31, 2015 to December 31, 2025. We divided the dataset into two subsets, using February 24, 2022 (the outbreak date of the Russia-Ukraine War) as the breakpoint. The first subset contains data from December 31, 2015, to February 23, 2022 (1548 days), and the second subset contains data from February 25, 2022, to December 31, 2025 (966 days). 

We applied the proposed test with $p=5$ to examine whether the directional dependence of OVX on XAU changed after February 24, 2022, and the p-value is 0.047, which indicates that the relationship between OVX and XAU had a significant structural change. 

To further demonstrate the change in the directional dependence of OVX on XAU, some cross-quantilogram heatmaps of OVX on XAU for the two periods (before/after the breakpoint) are presented in Figure 1. The lag lengths are set to 1, 5, and 22, corresponding to one day, one week, and one month, respectively. In all figures, red and blue indicate positive and negative cross-quantilogram values, respectively. The left and right panels of Figure 1 represent the pre- and post-Russia-Ukraine War periods, respectively. 

In the pre-war period, for each lag length, the current returns of XAU could be positively correlated with past OVX values when the current XAU returns are extremely high, as suggested by the large red region at the top of the heatmap. In addition, the current returns of XAU could be negatively correlated with past OVX values when the current XAU returns are extremely low, as suggested by the blue region at the bottom of each heatmap in the pre-war period. When the current XAU returns are not extreme, there is no clear pattern indicated by the heatmaps. In such case, it is possible that comparing to OVX, the XAU return was more influenced by other factors such as the US dollar exchange rate or interest rates (see Herley et al. \cite{Herley:2024} and Ndlovu and Ndlovu \cite{Ndlovu:2024}).

\begin{figure}[h]
    \centering
    
    \begin{subfigure}{0.35\textwidth}
        \includegraphics[width=4.2cm,height=4cm]{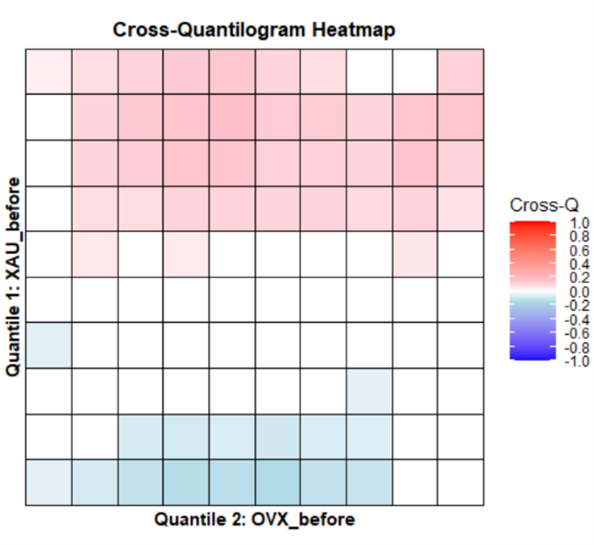}
        \caption{k=1}
    \end{subfigure}
    \begin{subfigure}{0.35\textwidth}
        \includegraphics[width=4.2cm,height=4cm]{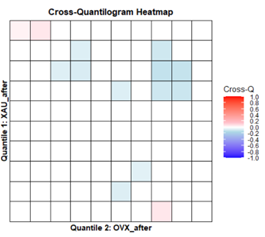}
        \caption{k=1}
    \end{subfigure}
    
    \vspace{0.5em} 
    
    \begin{subfigure}{0.35\textwidth}
        \includegraphics[width=4.2cm,height=4cm]{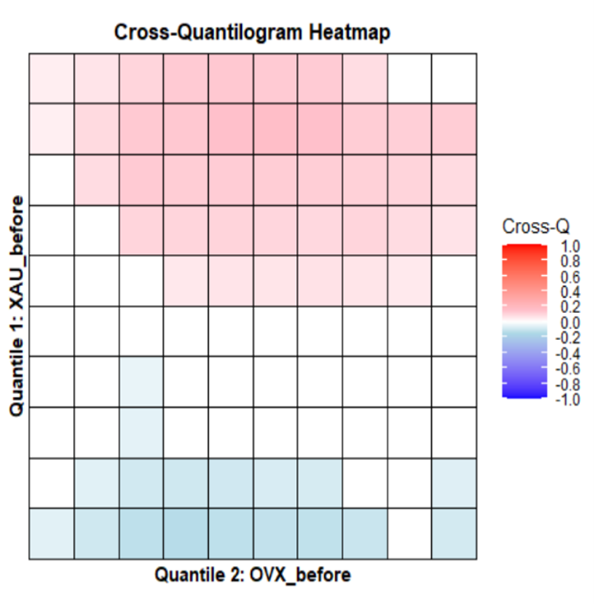}
        \caption{k=5}
    \end{subfigure}
    \begin{subfigure}{0.35\textwidth}
        \includegraphics[width=4.2cm,height=4cm]{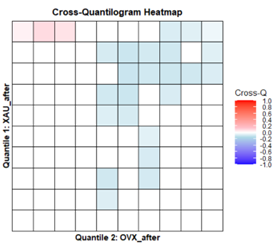}
        \caption{k=5}
    \end{subfigure}
  
   \vspace{0.5em} 
    
    \begin{subfigure}{0.35\textwidth}
        \includegraphics[width=4.2cm,height=4cm]{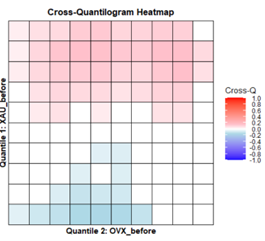}
        \caption{k=22}
    \end{subfigure}
    \begin{subfigure}{0.35\textwidth}
        \includegraphics[width=4.2cm,height=4cm]{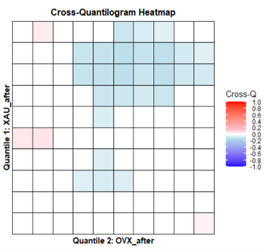}
        \caption{k=22}
    \end{subfigure}

    \caption{The cross-quantilogram heatmaps of OVX on XAU for the two periods}
\end{figure}

Comparing to the pre-Russia-Ukraine war period, the cross-quantilogram heatmaps
for the post-Russia-Ukraine war period exhibit a different pattern. Roughly speaking, some blue region appears in the upper-right corner of each heatmap after the outbreak of the war, and the blue region expands as the lag length increases. One possible explanation is that the increased volatility in oil prices after the outbreak of war led to gold being seen as a safe-haven asset, resulting in increased short-term demand for gold. Over time, profit-taking in the gold market has reduced the likelihood of having extremely high XAU returns. This phenomenon is consistent with the findings of Caporale and Plastun \cite{Caporale:2021}).

In the case where VIX was used as a controlling variable, the cross-quantilogram heatmap patterns of OVX on XAU returns are similar to those in the case where no controlling variable was used (see Figure 2). However, the strength of directional dependence of OVX on XAU returns became much weaker when VIX was used as a controlling variable. A possible explanation is that VIX can be highly correlated with OVX, in such case, most volatilities could be explained by VIX, which reduced the direction dependence of OVX on XAU returns. A similar point of view can be found in Wei et al. \cite{Wei:2024}, but the point in Wei et al. \cite{Wei:2024} was made with respect to green bond performance instead of XAU returns. 

\begin{figure}[h]
    \centering
    
    \begin{subfigure}{0.35\textwidth}
        \includegraphics[width=4.2cm,height=4cm]{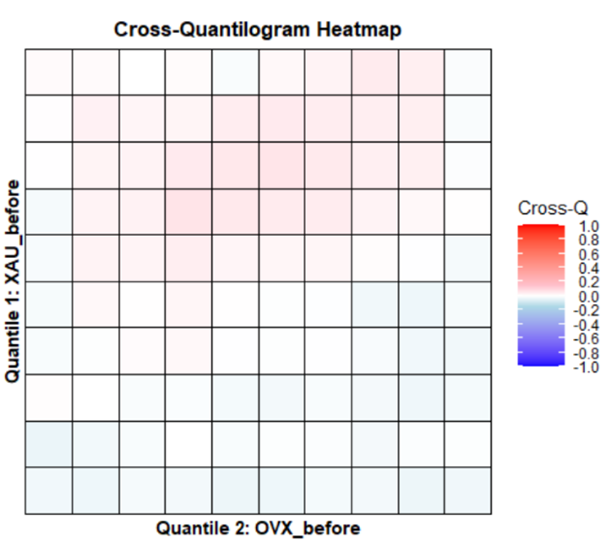}
        \caption{k=1}
    \end{subfigure}
    \begin{subfigure}{0.35\textwidth}
        \includegraphics[width=4.2cm,height=4cm]{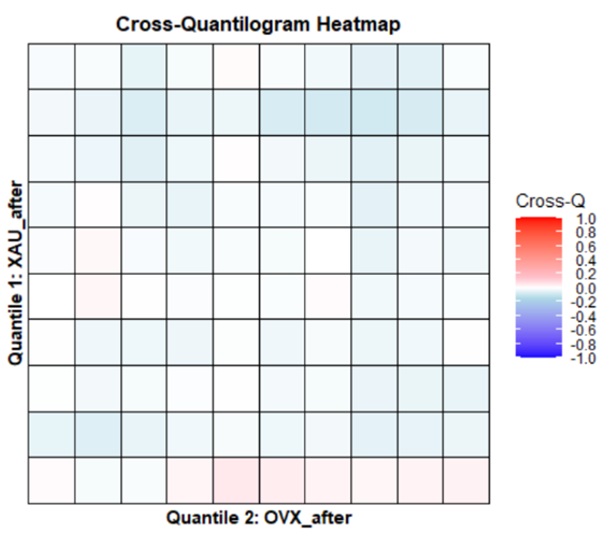}
        \caption{k=1}
    \end{subfigure}
    
    \vspace{0.5em} 
    
    \begin{subfigure}{0.35\textwidth}
        \includegraphics[width=4.2cm,height=4cm]{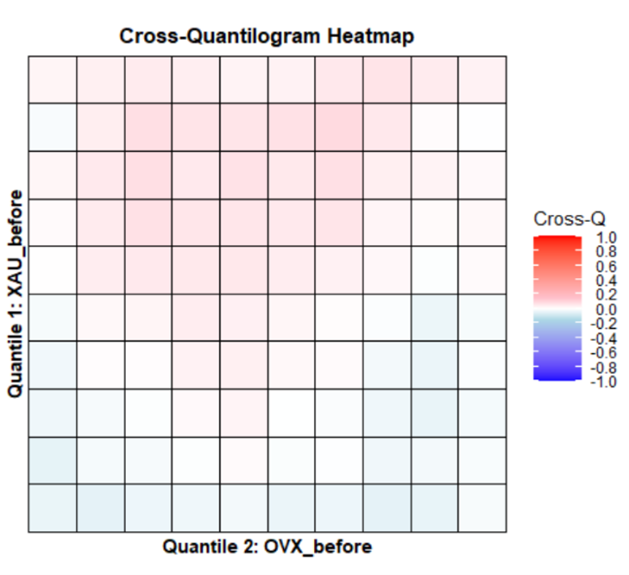}
        \caption{k=5}
    \end{subfigure}
    \begin{subfigure}{0.35\textwidth}
        \includegraphics[width=4.2cm,height=4cm]{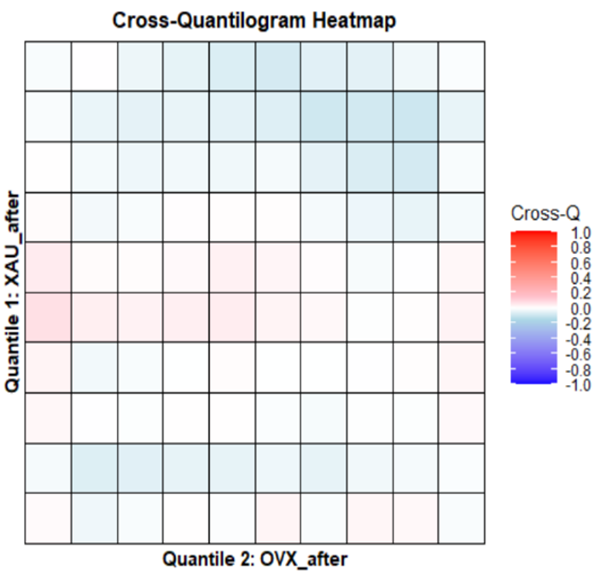}
        \caption{k=5}
    \end{subfigure}
  
   \vspace{0.5em} 
    
    \begin{subfigure}{0.35\textwidth}
        \includegraphics[width=4.2cm,height=4cm]{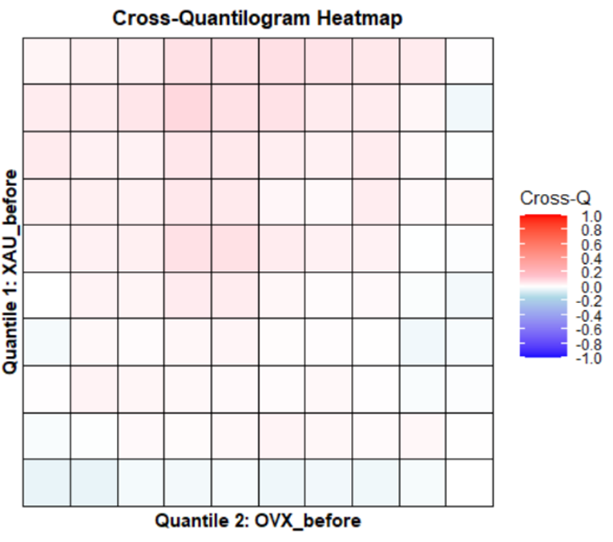}
        \caption{k=22}
    \end{subfigure}
    \begin{subfigure}{0.35\textwidth}
        \includegraphics[width=4.2cm,height=4cm]{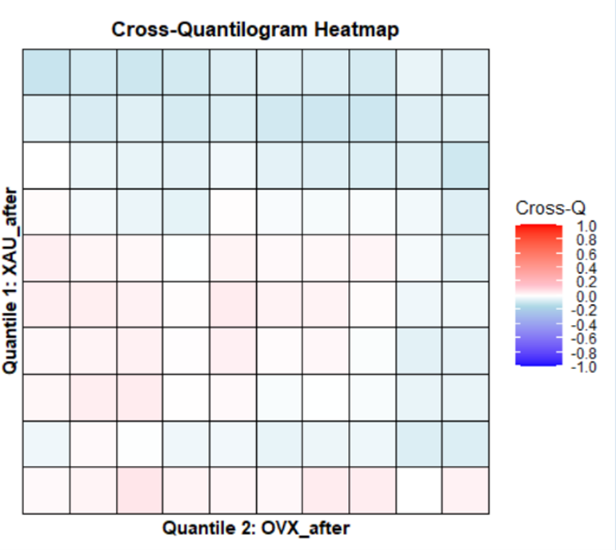}
        \caption{k=22}
    \end{subfigure}

    \caption{The cross-quantilogram heatmaps of OVX on XAU for the two periods with VIX as a controlling variable}
\end{figure}
 
\pagebreak
\section{Conclusion}
We proposed a test for testing differences in cross-quantilograms for two stationary series. Based on the results of our simulation study, the power performance of the proposed test is satisfactory when the differences  in  cross-quantilograms for the two series are large enough. 


\end{CJK}
\end{document}